\documentclass[conference]{IEEEtran}
\IEEEoverridecommandlockouts
\usepackage{cite}
\usepackage{amsmath,amssymb,amsfonts}
\usepackage{algorithmic}
\usepackage{graphicx}
\usepackage{textcomp}
\usepackage{xcolor}
\def\BibTeX{{\rm B\kern-.05em{\sc i\kern-.025em b}\kern-.08em
    T\kern-.1667em\lower.7ex\hbox{E}\kern-.125emX}}
    
\bstctlcite{IEEEexample:BSTcontrol}

\usepackage{lipsum}
\usepackage{xcolor}

\usepackage[hypertexnames=true]{hyperref}
\hypersetup{
	colorlinks,
	linkcolor={red!75!black},
	citecolor={blue!60!black},
	urlcolor={blue!80!black}
}

\def\BibTeX{{\rm B\kern-.05em{\sc i\kern-.025em b}\kern-.08em
    T\kern-.1667em\lower.7ex\hbox{E}\kern-.125emX}}

\usepackage{booktabs}
\usepackage{multirow}

\begin{document}
\bstctlcite{IEEEexample:BSTcontrol}
\title{A CMOS-based Characterisation Platform for Emerging RRAM Technologies
}

\author{
    \IEEEauthorblockN{Andrea Mifsud$^{*\dagger}$, Jiawei Shen$^{*}$, Peilong Feng$^{*\dagger}$, Lijie Xie$^{*}$, Chaohan Wang$^{*}$, Yihan Pan$^{\P}$,  Sachin Maheshwari$^{\P}$, \\  Shady Agwa$^{\P}$, Spyros Stathopoulos$^{\P}$, Shiwei Wang$^{\P}$,  Alexander Serb$^{\P}$, Christos Papavassiliou$^{*}$, \\Themis Prodromakis$^{\P}$,  Timothy G. Constandinou$^{*\dagger\ddag}$}
    \IEEEauthorblockA{$^*$Department of Electrical and Electronic Engineering, Imperial College London, SW7 2BT, UK\\$^\dagger$Centre for Bio-Inspired Technology, Institute of Biomedical Engineering, Imperial College London, SW7 2AZ, UK\\$^\ddag$Care Research \& Technology Centre, UK Dementia Research Institute, UK\\$^\P$Centre for Electronics Frontiers, Electronics and Computer Science, University of Southampton, SO17 1BJ, UK\\
    Corresponding author email: a.mifsud@imperial.ac.uk
    }%
}%

\maketitle

\begin{abstract}
Mass characterisation of emerging memory devices is an essential step in modelling their behaviour for integration within a standard design flow for existing integrated circuit designers. This work develops a novel characterisation platform for emerging resistive devices with a capacity of up to 1 million devices on-chip. Split into four independent sub-arrays, it contains on-chip column-parallel DACs for fast voltage programming of the DUT. On-chip readout circuits with ADCs are also available for fast read operations covering 5-decades of input current (20\,nA to 2\,mA). This allows a device's resistance range to be between 1\,k$\Omega$ and 10\,M$\Omega$ with a minimum voltage range of $\pm$1.5\,V on the device.
\end{abstract}



\section{Introduction} 
\label{s:intro}

Emerging memory technologies including resistive random access memory (ReRAM/RRAM)~\cite{jain_36mb_2019,fackenthal_16gb_2014,kawahara_8_2013,liu_1307mm2_2013} provide potential solutions to the challenges faced by current memories due to their lower power consumption, scalability, non-volatility and high-speed operation. Apart from the typical use case as computer memory, such devices have other uses in applications such as in-memory computing \cite{xue_embedded_2020,chi_prime_2016} and FPGAs \cite{liauw_nonvolatile_2012,dao_memristor-based_2021}. A common criteria of these applications is the use of large arrays with millions, billions or more of devices in a single chip \cite{antoniadis_open-source_2021}. Such scale brings additional challenges to these emerging technologies which need to be solved before reaching their full potential.

One such challenge is variability of a device's state as the interfacing periphery circuitry should satisfy the entire working range to ensure optimal operation. Another challenge is the yield of the device and/or integration process~\cite{irds_international_2020,meena_overview_2014}. Being able to measure, model \cite{maheshwari_design_2021} and optimise this as per the application is vital in prolonging the life-span of the product.

Several of these emerging memory technologies are two terminal devices that operate by varying their port-to-port resistance~\cite{irds_international_2020,irds_international_2020-1}. Varying this resistance is enabled by applying a voltage/current pulse/s within a set range, whilst reading the resistance is executed by using conventional methods; apply current and read voltage drop or vice-versa. Thus whilst characterising a few devices is possible using bench-top equipment~\cite{tektronix_pulse_nodate}, mass characterisation of devices requires a custom PCB~\cite{lupo_custom_2021,cayo_design_2021,foster_fpga_2020,fuente_development_2020} or chip-based platform~\cite{maeda_resistance_2020}. Whilst a PCB-based platform is easy to customise, it limits the operation and number of devices under test (DUTs) due to the parasitic resistance and capacitance connected to the devices. On the other hand, a chip-based platform minimises the parasitics, enabling higher speed of operation whilst the DUTs are integrated within an environment that is also typical of the application. \\

This paper presents the top-level architecture to a novel chip-based characterisation platform for emerging resistive memory technologies with the capability of testing up to 1 million devices integrated on top of a 180\,nm CMOS process. Section~\ref{s:sysAr} describes the system architecture and design implementation. Section~\ref{s:integration} then covers the DUT - CMOS integration process step for the case of RRAM. Section~\ref{s:results} presents simulated results and Section~\ref{s:conclusion} concludes the work.

\begin{figure}[!t]
\centering
\includegraphics[width=\linewidth]{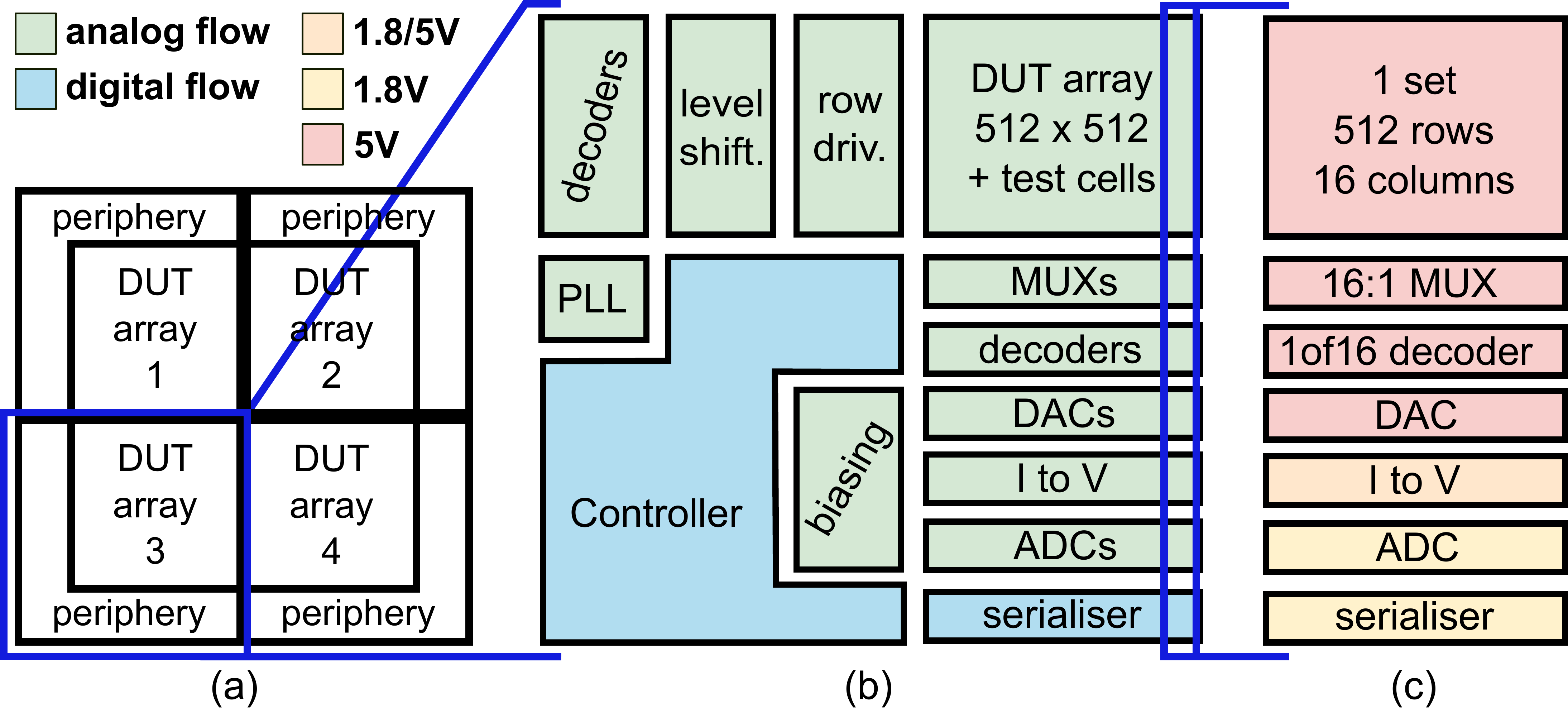}
\caption{System architecture: (a) Top level view; (b) array focused; (c) set focused.} 
\label{f:sysArch}
\end{figure}

\section{System Architecture}
\label{s:sysAr}
Fig.~\ref{f:sysArch} shows a simplified block diagram of the proposed characterisation platform. The platform consists of 4 identical and independently controlled sub-arrays, each with 512 rows and columns, bringing the total to 1 million 1T1R cells on chip. Due to layout constraints, 16 columns (also referred to as a set) share the same column-parallel circuitry which consists of a DAC, a current-to-voltage (I to V) converter, and an ADC. This enables 32 devices to be characterised in parallel from each sub-array at any time, with a total of 128 devices when all sub-arrays are operational. All data from the 32 sets is then sent off-chip through a serialiser. System control is done through the on-chip controller fully configured through the SPI interface.

\subsection{1T1R Cell}
\begin{figure}[!t]
\centering
\includegraphics[width=\linewidth]{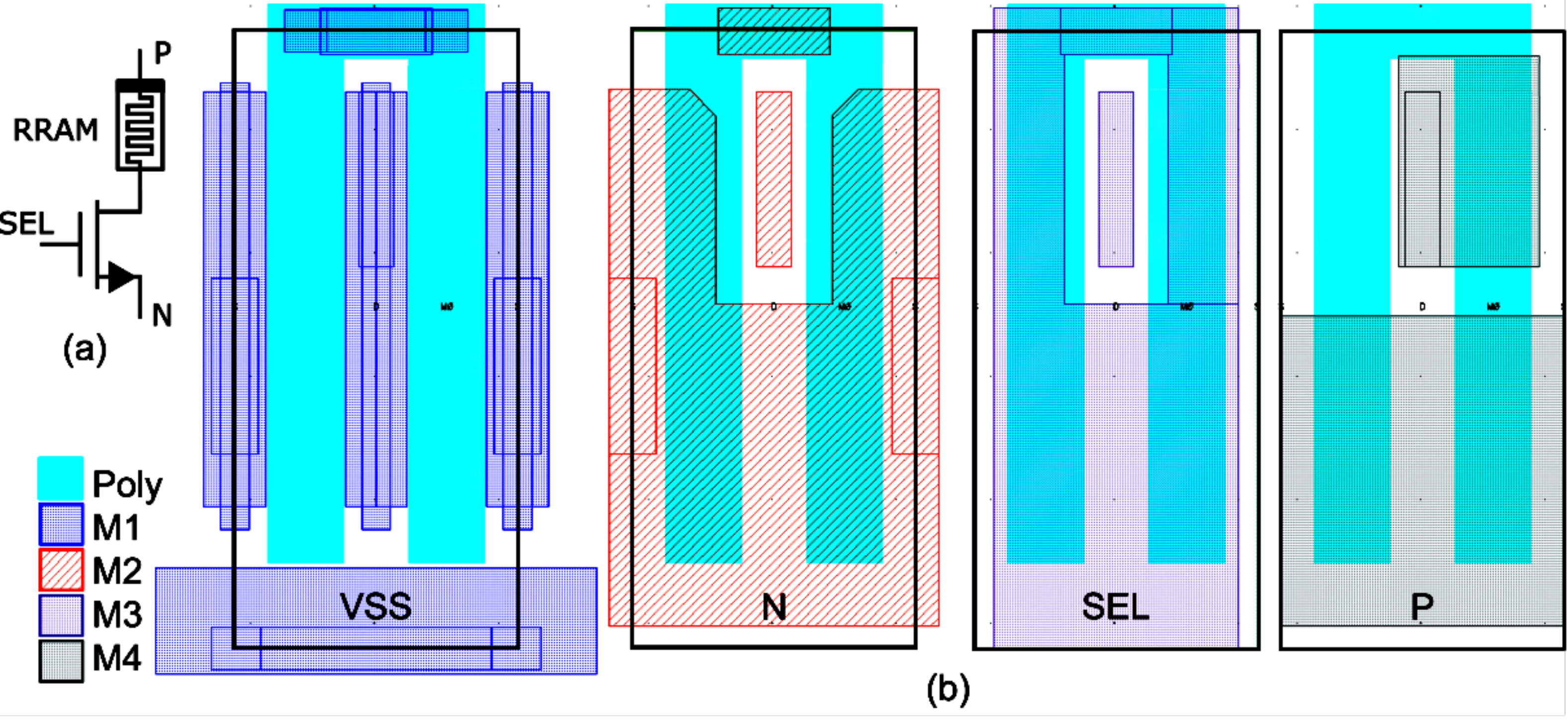}
\caption{1T1R cell; (a) schematic using a nMOS transistor with W/L of 7.4\,$\mu$m/0.6\,$\mu$m; (b) layout of cell and metal layers above the transistor; M2 for net \textit{N}, M3 for net \textit{SEL} and M4 for net \textit{P} and intermediate DUT-nMOS net.} 
\label{f:pixSch_lyt}
\end{figure}

The standard 1T1R structure (Fig.~\ref{f:pixSch_lyt}(a)) is designed with a 5\,V nMOS transistor and a RRAM whose minimum resistance is assumed to be 1\,k$\Omega$. It allows writing the RRAM in both forward and reversed direction by setting terminal P/N to a specific voltage up to 5\,V and the other terminal N/P to 0\,V. To satisfy the RRAM writing scheme, 1.5\,V in both directions is required to change the device's state. As a result, an nMOS transistor is chosen due to a higher mobility than a same sized pMOS, thus minimising cell footprint. For the layout in Fig.~\ref{f:pixSch_lyt}(b), the RRAM is placed above the transistor to maximise the cell density as explained further in Section~\ref{f:dut-integration} resulting in cell dimensions of 5\,$\mu$m $\times$ 2.28\,$\mu$m, which are mainly determined by the size of the transistor. Each signal path is extended to cover as much area as possible to reduce the parasitic resistance of the track and enable faster settling times throughout the array.

\subsection{Row Circuits}
The row circuits consist of digital circuits to select one or more rows at a given time. Selecting just one cell is possible through the use of an address decoder which takes the gray-coded address provided by the controller and enables the corresponding row. Once a row is enabled, an independent \textit{select} signal is then used to control the duration of the pulse. This gives finer control over the pulse width.

\subsection{Column Circuits}
The objective of the column circuits is two-fold; (1) apply a voltage across a DUT, and (2) measure the current flowing through it. This is possible through the circuits described below.

\subsubsection{Column Switches and Decoder}
Due to layout constraints a number of columns share the same set circuitry. As a result, analog switches are used to route the appropriate column to the circuit in use. The first level implements an analog multiplexer to enable one column out of the available 16 in a set. These switches are controlled by a decoder that takes a 4-bit binary input and enables a single column. The second level determines the polarity of the voltage applied to the \textit{P} and \textit{N} lines of the selected column. This is very similar to an H-bridge configuration, with the column of interest being the driven load.

\subsubsection{DAC}
This DAC generates the specified voltage to drive the DUT array to perform the desirable sweeps or characterisation procedures. The challenge for this DAC design is to drive a wide load range from 1\,k$\Omega$ to 5\,M$\Omega$. Also, its desired voltage output range is defined as 50\,mV to 4\,V (range decreases for low resistance).  A traditional 8-bit voltage-mode R-2R DAC is implemented with 5\,V MOSFETs to achieve an output range from 50\,mV to 4.9\,V. However, the final output range is determined by a high-drive strength rail-to-rail output buffer. This buffer uses a conventional complementary rail-to-rail amplifier together with a constant-gm circuit~\cite{Shouli2005_constantgm} as the input stage. Also, the indirect compensation technique~\cite{Saxena2006_Indirect} is used at the output stage to ensure a good phase margin and high unity-gain frequency for a wide load range. 

\subsubsection{Current-To-Voltage Converter}

\begin{figure}[!t]
\centering
\includegraphics[width=0.8\linewidth]{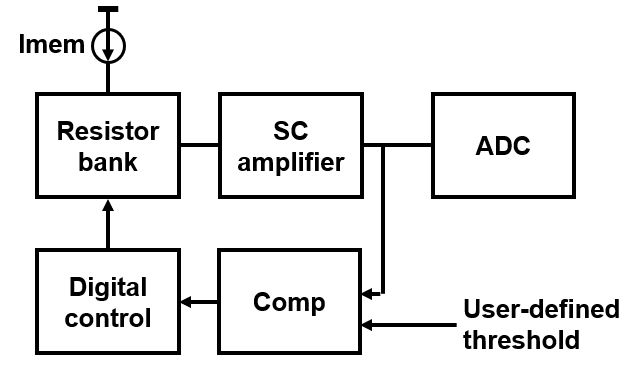}
\caption{A block diagram of the proposed readout system.} 
\label{f:readout_system}
\end{figure}

To enable an on-chip and fast characterisation, an on-chip readout system is implemented of which a block-level overview is depicted in Fig.~\ref{f:readout_system}. By configuring the DAC, a user-defined reading voltage is applied to the DUT, that generates a specific current based on resistance of the DUT that flows from the DAC, through the DUT and finally through the current-to-voltage converter. This current signal is captured, converted and amplified in the proposed readout system. 

A 5-stage logarithmic resistor bank is employed to convert the input current to a voltage-mode signal for amplification. A switched-capacitor (SC) amplifier is then used to amplify the small output voltage of the resistor bank to ensure proper use of the full input range of the ADC. To keep the voltage across the resistor bank within a small range (less then 50\,mV), a comparator with a programmable threshold is used. Additionally this threshold determines the voltage range at the output of the amplifier which is then fed to the ADC.
The algorithm used for the resistor bank control is very similar to that found in an SAR ADC. The resistor with the lowest resistance is activated during the first clock cycle and the amplified output is compared with the threshold. If the output is higher, the algorithm terminates here. In the other instance, the algorithm moves to the next resistor and repeats the process until the amplified output is either larger than the threshold, or the last resistor has been selected. When terminated the state of the selected resistor is stored and sent to the serialiser for off-chip data transfer.

\subsubsection{ADC}
A 12-bit SAR ADC digitises the resulting voltage from the preceding I-to-V converter block. It has been designed to operate with a sampling frequency of 250\,kHz and an input range 0.1\,V to 1.7\,V. The two major challenges in this design have been: (1) minimising power consumption and (2) 12-bit resolution without on-chip calibration. This ADC consists of shift registers for timing signal generation, a capacitor type DAC, and a dynamic comparator. The DAC employs a fully differential C-2C capacitor array to minimise effects due to parasitic capacitance. It directly uses the bottom plate of capacitor arrays to sample and hold the input signal. A dynamic comparator~\cite{razavi_strongarm_2015} is used for lower kick-back noise and lower energy consumption. To minimise the area used by the capacitor arrays, the minimum unit capacitor is used for the DAC, resulting in a total area of 560\,$\mu$m $\times$ 60\,$\mu$m for the ADC. In addition, the capacitor used for the MSB check is split into a copy of the remaining capacitors~\cite{ginsburg_energy-efficient_2005} to minimise power consumption and obtain a faster charge/discharge transient at the cost of a higher number of control signals. 

\subsection{Serialiser}
A two-stage serialiser is added at the bottom of the ADC stage to capture the data generated by the 32 sets (16 columns each) in addition to two extra control packets. Every set provides 26-bits of data, 12-bits of which come from the ADC, another 5-bits from the I-to-V converter and the remaining bits are added to the packet to provide further information about the column including its address. The two control packets also provide feedback on the running operations. The main idea of the two-stage serialiser is to use the first stage to browse the packets quickly and then use the second stage to serialize the targeted packet with 2 bits out of the chip per cycle. The first stage is a 34-entry shift-register with 26-bit resolution which captures the 32 data packets from the 32 sets in addition to the 2 control packets.  This big shift-register browses the 34 packets within the minimum number of clock cycles and delivers the targeted 26-bit packet to the second stage which is a small 26-bit shift-register. The small shift-register serialises the data out of the chip with 2 bits at a time to increase the overall transmission throughput whilst minimising lane data rate. Unlike a one-stage serialiser, the two-stage design serialises the packet number N completely out after (1+N+13) cycles instead of  (N$\times$13 + 13) cycles. The serialiser is implemented using the digital ASIC flow.

\subsection{System controller}
All the blocks mentioned earlier require a number of control signals for operation. A controller has been designed to generate these signals based on the current system configuration that is written through a serial interface. The controller supports two main operations; (1) a write and (2) a read. During a write, the appropriate row and column/s are selected, the DAC voltage set, and the cell enabled for a specific pulse width. This operation is used to program the device without measuring the resulting current. In the case of RRAM, it is expected that during a write, given appropriate values for the voltage and pulse width, the device will experience a change in resistance. On the other hand, apart from applying a voltage, a read operation also enables the I-to-V converter, ADC and serialiser to measure the current, digitise it, and transfer the data off-chip. In the case of RRAM, this operation is used to run the conventional IV sweep on the device, or to measure the current resistive state of the DUT (e.g. $V_{read} = $ 0.5\,V).
\begin{figure}
  \centering
  \includegraphics[width=0.9\linewidth]{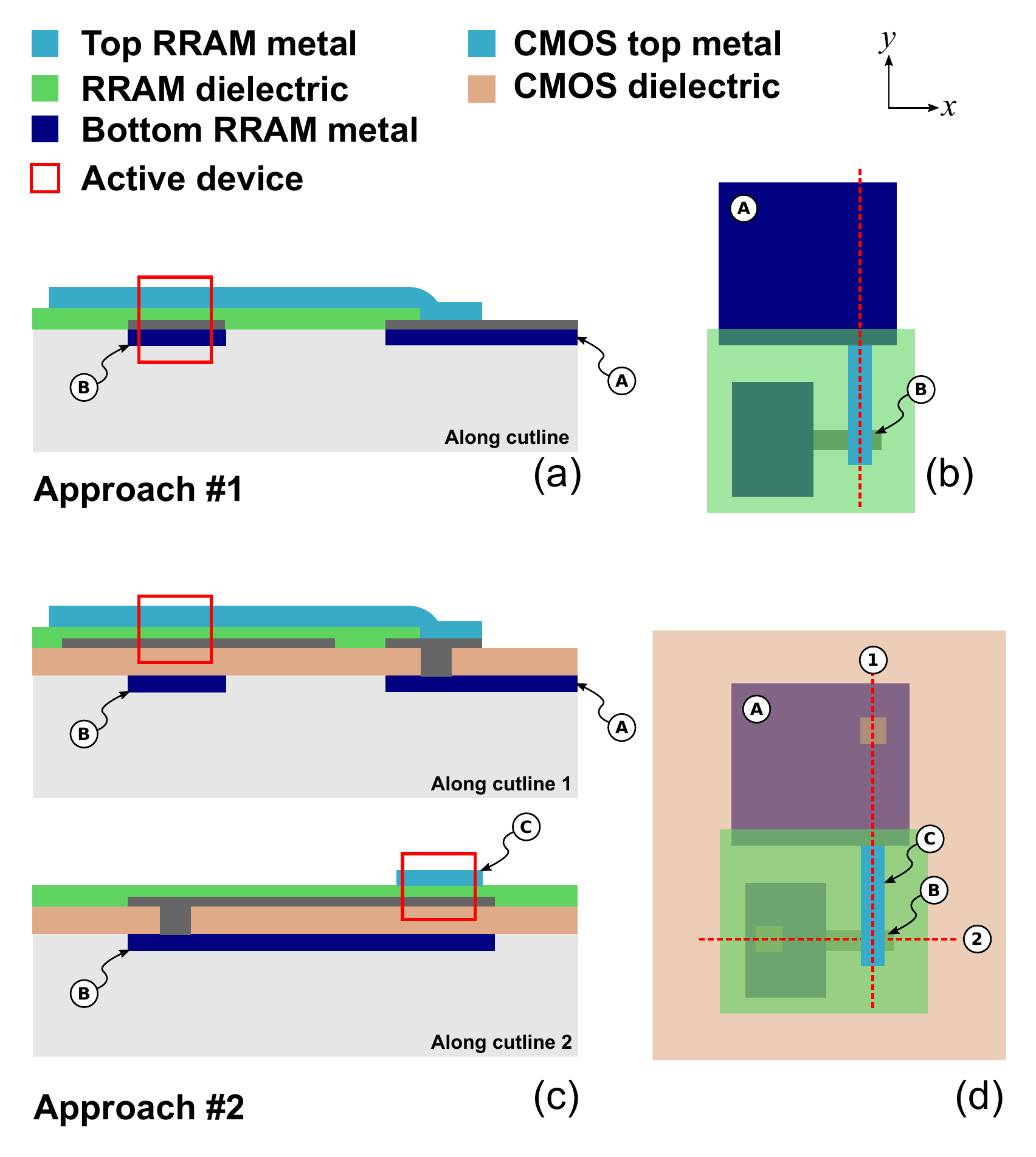}
  \caption{\label{f:dut-integration}CMOS-DUT integration approaches
  as described in Section~\ref{s:integration}. In the via-less approach (a, b)
  RRAM is connected directly to the top-most CMOS metal whereas
  in the second approach (c, d) this is done through vias in the CMOS
  capping dielectric.}
\end{figure}

\section{DUT-CMOS Integration}
\label{s:integration}

The proposed platform is oriented towards the characterisation of two terminal
DUTs. Integrating the actual devices onto it is done using regular
back-end-of-line processes (BEOL) after the CMOS substrate has been formed.
There are two approaches explored, depending on the way the CMOS layer
interfaces the DUT.\@ The first involves etching away the capping dielectric
and fully exposing the final CMOS metal layers. This approach presents a
simpler integration route as no vias are required to connect to the circuitry
below the device. It also eliminates any additional parasitic capacitance
that may be present between the bottom RRAM electrode and the top-most CMOS
metal layer. This comes at the cost of increased routing complexity on the BEOL
layers. The second approach retains the capping dielectric and uses vias to
connect to the circuitry below. This simplifies routing complexity (as the CMOS
and RRAM layers are fully separate) at the cost of increased processing
complexity (vias and additional parasitic capacitance).
Fig.~\ref{f:dut-integration} summarises the two approaches.

The prototype devices consist of a Metal-Insulator-Metal structure (MIM)
in the form of BE/AL/TE (bottom electrode, active layer and top electrode
respectively). Electrodes (BE/TE) can be a selection of platinum (Pt), gold
(Au) and silver (Ag) for different types of interfacial behaviour 
(Schottky for Pt, ohmic for Au~\cite{Michalas2018} and diffusive/filamentary for Ag) with nominal thicknesses of 20/30/50\,nm respectively. AL materials that have
been considered are TiO$_2$, AlO$_\text{x}$/TiO$_2$~\cite{Stathopoulos_2017} or
ZnO~\cite{Simanjutak2021}. All RRAM materials are deposited using RF magnetron
sputtering in either an argon atmosphere (for Pt) or oxygen/argon atmosphere. Au
deposition has been performed using electron beam evaporation. All layers are
defined with e-beam lithography (using 200/200\,nm PMMA/MMA resist) and lift-off
in NMP bath. Capping dielectric is stripped using ion-beam milling either fully
(in the first, via-less, approach) or partially (in the second approach).
\begin{figure}[!t]
\centering
\includegraphics[width=\linewidth]{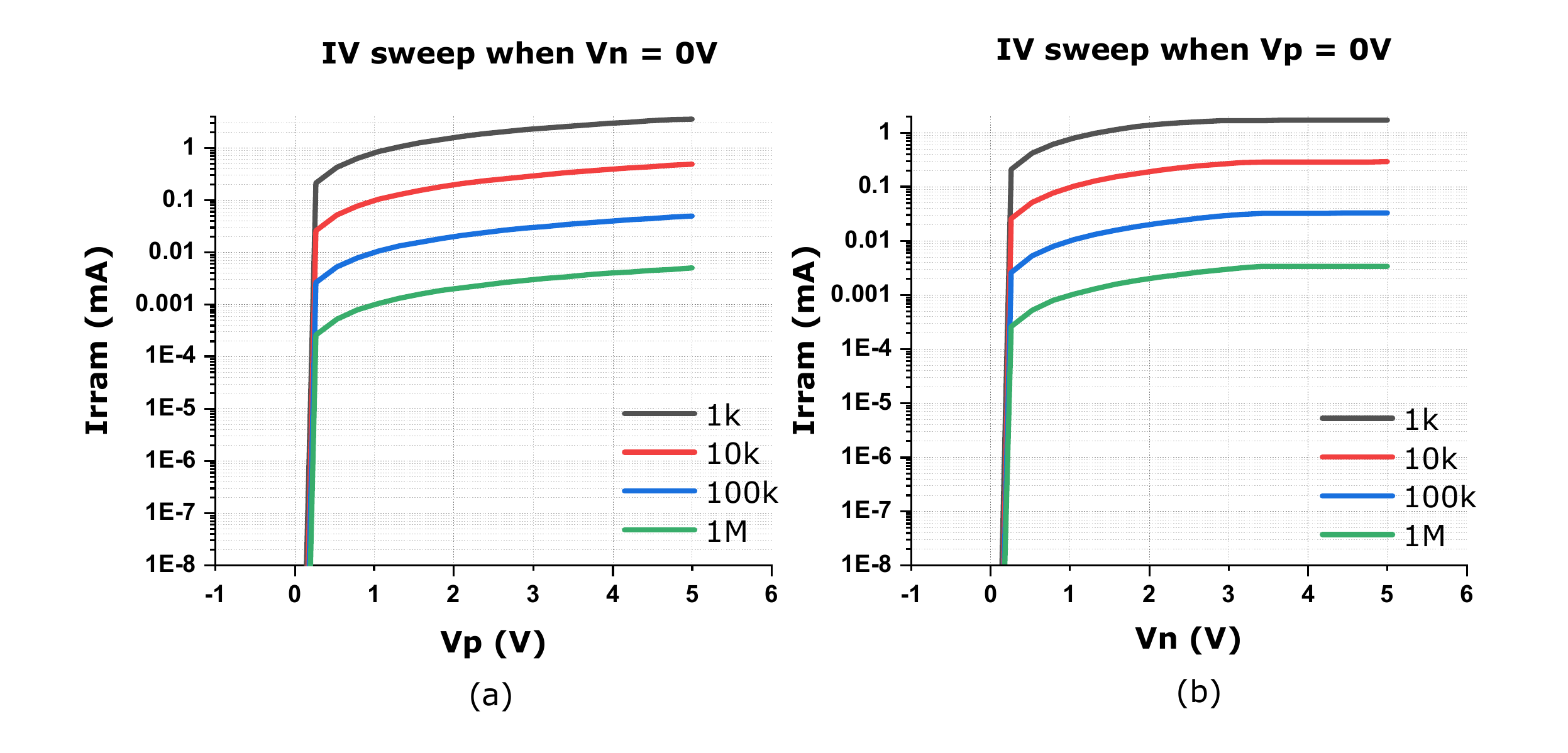}
\caption{IV plots of the cell at different resistance values; (a) forward mode (+ve V across DUT) and (b) reversed direction (-ve V across DUT).} 
\label{f:1t1rIVsweep}
\end{figure}

\begin{figure}[!t]
\centering
\includegraphics[width=\linewidth]{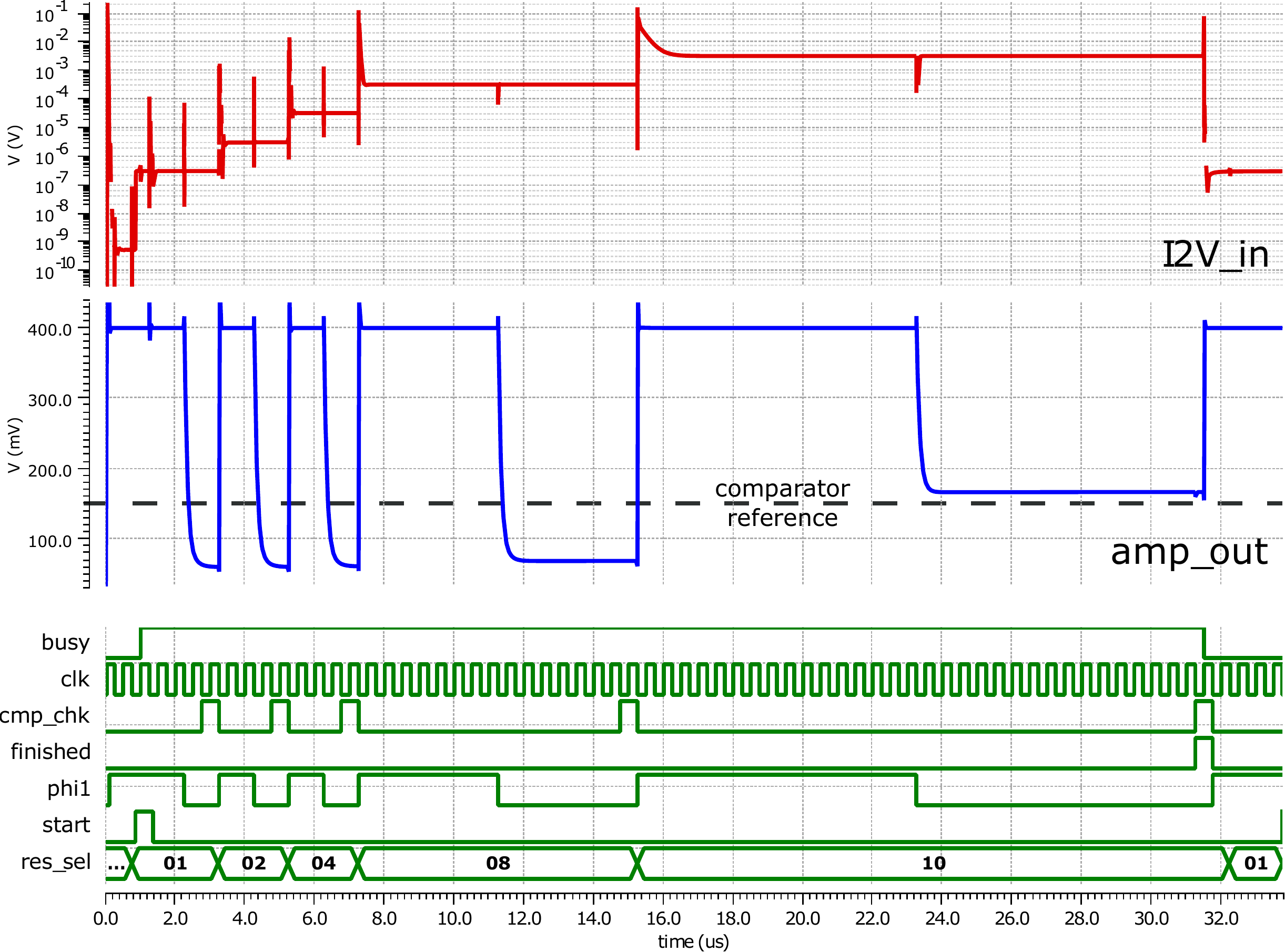}
\caption{Operation of I-to-V circuit for a 20\,nA input current; \textit{I2V\_in} is the input node; \textit{amp\_out} is the output of the amplifier; \textit{cmp\_chk} is the point of comparison to the voltage reference; \textit{res\_sel} is the resistor select (0x01 - least resistance to 0x10 - highest resistance).} 
\label{f:i2v_tran}
\end{figure}

\begin{figure}[!t]
\centering
\includegraphics[width=\linewidth]{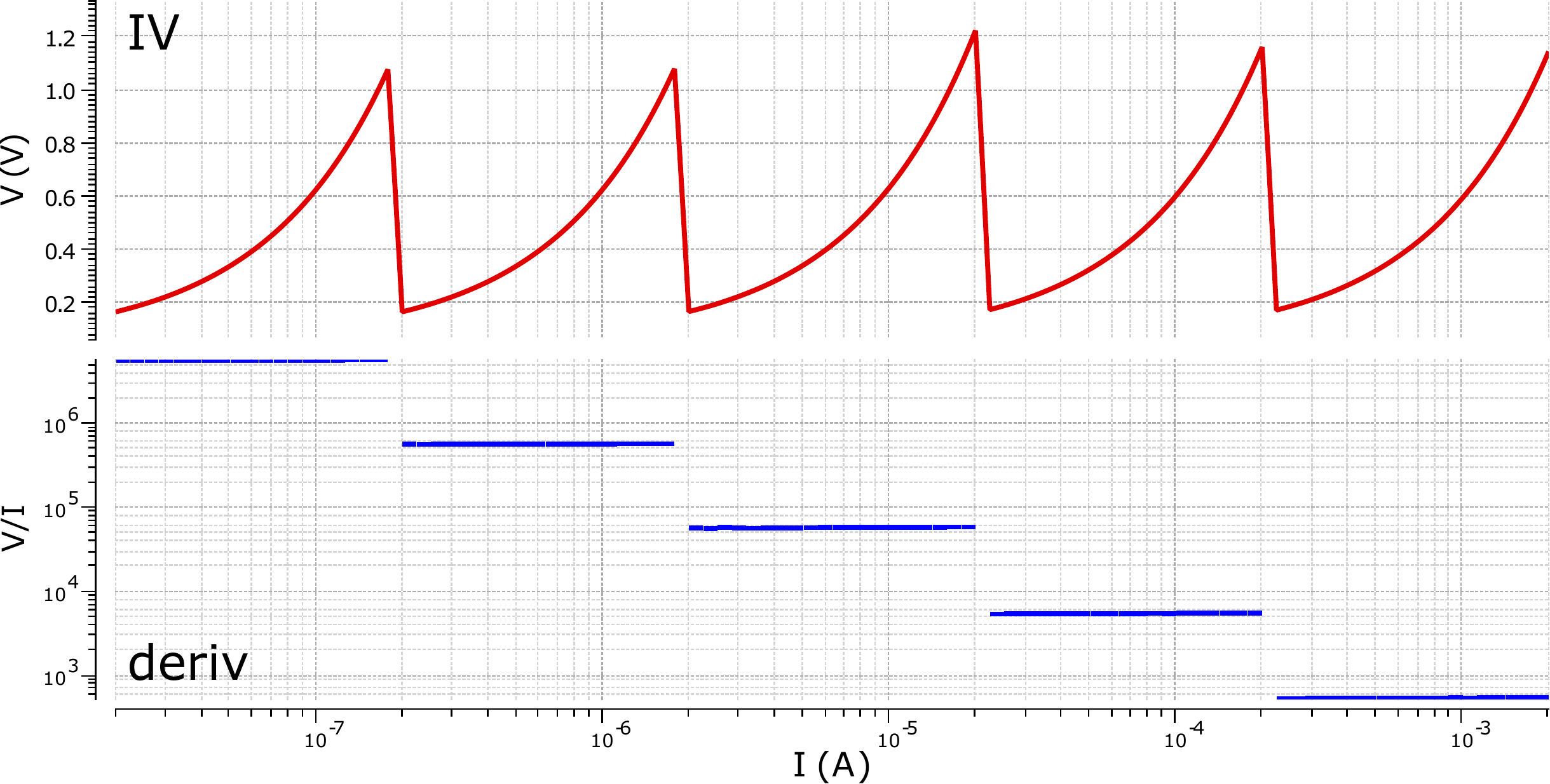}
\caption{IV plot of I-to-V circuit. Points shown are for the final selected resistor for that conversion; \textit{deriv} is the derivative showing the five different gains.} 
\label{f:i2v_iv}
\end{figure}

\section{Results and Discussion}
\label{s:results}
The 1T1R cell was simulated for multiple resistor values, mimicking the effect of a DUT under the same condition. The resulting current range is shown in Fig.~\ref{f:1t1rIVsweep}. The maximum current supported through is approximately 1.5\,mA (with 1.5\,V across a 1\,k$\Omega$ DUT). On the other hand, given the range of resistance and voltage used, the resulting current range spans 5 decades, which match the 5 stages in the I-to-V converter.

A transient simulation of this circuit is shown in Fig.~\ref{f:i2v_tran} with an input current equal to 20\,nA. With the use of a switched-capacitor amplifier, there are two phases for each resistor; a sampling (and resetting) phase and an amplifying phase. Additionally as evinced in Fig.~\ref{f:i2v_tran}, both the sampling and the amplifying phase for the highest resistance are longer when compared to that of the lowest resistance in the resistor bank. This is there to optimise the time per stage as required by the settling time analysis of the entire set circuit.
A full characterisation of the circuit is illustrated in Fig.~\ref{f:i2v_iv}. This shows the nominal behaviour of the circuit over a wide range of current. Additional specifications for the chip are listed in Table~\ref{t:specs}.

\begin{table}[t]
\centering
  \caption{System Specifications}
  \footnotesize
    \begin{tabular}{ll}
    \toprule
    Parameters & Value \\
    \midrule
    Guaranteed max. V across DUT & $\pm$1.5\,V \\
    DUT resistance range &  1\,k$\Omega$ - 10M\,$\Omega$ \\
    \midrule
	\# of cells, \# of test cells & 1M, 32k \\
     Cell configuration, cell size & 1T1R 5\,V, 5um x 2.28um \\ 
    Min. pulse width & 5ns \\
    \midrule
    DAC resolution, output range & 8\,bits, 50\,mV - 3\,V \\
    \midrule
    Current measurement range & 20\,nA - 2\,mA \\
    ADC resolution, sample rate & 12\,bits, 250\,kSPS\\
    \bottomrule
    \end{tabular}%
  \label{t:specs}%
\end{table}%
\section{Conclusion}
\label{s:conclusion}

This work presents a CMOS-based, on-chip characterisation platform for emerging memory technologies with particular focus on forming-free RRAM. The chip consists of four independent sub-arrays with a full resolution of one million devices. On-chip DACs provide a programmable voltage across the device, whilst current-to-voltage and ADC circuits are used to digitise the measured current over a 20\,nA to 2\,mA range.

\section*{Acknowledgment} 
The authors acknowledge the support of the EPSRC FORTE Programme Grant (EP/R024642/1), the RAEng Chair in Emerging Technologies (CiET1819/2/93), as well as the EU projects SYNCH (824162) and CHIST-ERA net SMALL.


\bibliographystyle{IEEEtran}
\bibliography{IEEEabrv,refs.bib}

\end{document}